\documentstyle{article}
\begin{document}
\title{
Asymptotical small-$x$ behaviour of the spin structure
functions $g_1$ and $g_2$.
}

\author{
B.I.Ermolaev\thanks{Supported in part by the Grant
INTAS-RFBR-95-311}\\
{\it Ioffe Physico-Technical
Institute, St.Petersburg 194021, Russia}\\
S.I.Troyan\thanks{Supported in part by
the Grant R2G 300 from International Science Foundation}\\
{\it Petersburg Nuclear Physics Institute,
Gatchina 188350, Russia}}

\date{}

\maketitle

\begin{abstract}
We show that for small $x$ and in the double logarithmic approximation
(DLA) $g_2$ can be expressed through derivative of $g_1$ with respect to
logarithm of the QCD coupling. Therefore the small-$x$ behavior of
the both structure functions is similar. The analytical expression for
flavor nonsinglet structure function $g_1$ is presented and its
asymptotical behaviour is calculated.
\end{abstract}

In framework of the conventional parton model the
process of DIS can be considered as a superposition of two
independent phases:  hadron fragmentation in  constituent partons and
further DIS of a lepton on one of these partons.
At asymptotically high energies, $s\to\infty$, the behaviour of
the cross section is governed by the second phase.

The  spin-dependent part of the hadronic tensor for the second phase
with the initial parton being a quark, $W_{[\mu\nu]}$, can be written
in terms of
longitudinal and transversal projections of the quark spin,
relative to the plane of the incoming 4-vectors $p$ (quark) and $q$
(virtual photon), in a following way:
\begin{eqnarray}\label{structures}
& &W_{[\mu \nu]} = i\epsilon_{\mu\nu\alpha\beta} \frac{m
q_{\alpha}}{pq} \left( g_{\|}{\cal S}^{\|}_{\beta} +
g_{\perp}{\cal S}^{\perp}_{\beta} \right) \\
& &g_{\|} = g_1,\qquad g_{\perp} = g_1 + g_2 \nonumber
\end{eqnarray}
where $m$ and $\cal{S}$ are mass and spin of the quark,
and $g_1$, $g_2$ are functions of $x=Q^2/2pq$ and  $Q^2= -q^2>0$.

In the Born approximation
\begin{equation}
\label{born1}
g_{\perp} = g_{\|} = \frac{e^2_q}{2} \delta(1-x),
\end{equation}
with $e_q$ being the electric charge of the quark. This means that
in the Born approximation $g_2=0$.

In higher QCD orders at $\ln(1/x) \gg \ln(Q^2/\mu^2)$
all contributions of order of $(\alpha_s(\ln x)^2))^k$ must be taken
into full account.

$g_1$ was calculated in\cite{BER1,BER2} in the DLA analytically for
flavour nonsinglet (NS) case and numerically for singlet (S).
Here we present the analytical expression for $g_1^S$ and the new
result\cite{ET2157} that the structure function
$g_2$ can be
obtained through differentiation in QCD coupling $\alpha_s$ the analytical
expression for $g_1$.

The nonsinglet quark ladder diagram with $n$ gluon rungs
in Feynman gauge gives the following contribution to the
DLA for antisymmetric structure function
\begin{equation}\label{DLladd}
W_{\mu\nu}^{(n)}=\prod_{i=1}^n \left[\!\frac{\alpha_s C_F}{4\pi}
\int_x^{\beta_{i-1}}\!\frac{d\beta_i}{\beta_i}
\int_0^{2\pi}\!\frac{d\phi_i}{2\pi}
\int_{\kappa_i}^{s\beta_i}\!\frac{dk_{i\perp}^2}{[k_i^2]^2}
\right] \beta_n\frac{e_q^2}{2s}
\delta(\beta_n\!-\! x\!-\!\frac{\vec{k}_{n\perp}^2}{s}) T_{\mu\nu}^{(n)} ,
\end{equation}
with
$$\kappa_i = \mu^2+\frac{\beta_i}{\beta_{i-1}}\vec{k}_{i-1 \perp}^2 $$
where $k_i=\alpha_i q\!+\!\beta_i p\!+\!k_i^{\perp}$ is 4-momentum of
the quark in a cell $i$, and
\begin{equation}\label{split}
T_{[\mu\nu]}^{(n)} =
i\varepsilon_{\mu\nu\lambda\sigma} q^{\lambda} \frac{1}{2}{\em
Tr}\left\{ (\hat{p}\!+\!m)(1\!-\!\gamma_5\hat{\cal S})
\gamma_5\Gamma^{\sigma} \right\},
\end{equation}
where
\begin{equation}\label{biggam}
\Gamma^{\sigma} = \left(\prod_{i=1}^n
[-g^{\lambda_i{\lambda_i}^{\prime}}] \right)
\gamma_{\lambda_1^\prime}\hat{k_1}\ldots\gamma_{\lambda_n^\prime}\hat{k_n} \gamma^{\sigma}
\hat{k_n}\gamma_{\lambda_n}\ldots\hat{k_1}\gamma_{\lambda_1}.
\end{equation}

The key point for understanding the structure of the matrix
$\Gamma_{\sigma}$ is that
\begin{equation}\label{ftensor}
\hat{k}\gamma^{\sigma}\hat{k} = - k^2 f^{\sigma\sigma^\prime}(k)
\gamma_{\sigma^\prime} , \qquad
f^{\sigma\sigma^\prime}(k) = g^{\sigma\sigma^\prime} -
2\, \frac{k^{\sigma} k^{\sigma^\prime}}{k^2} .
\end{equation}

This means that $\Gamma_{\sigma}$ is linear in $\gamma^{\sigma}$ and
can be expressed in terms of convolution of $n$ similar $f$-tensors:
\begin{equation}\label{azimut}
\Gamma^{\sigma} = \left(\prod_{i=1}^n \left[ - 2k_i^2 \right]\right)
{\cal E}^{\sigma}_{\tau}\gamma^{\tau} , \qquad
{\cal E}_{\sigma\tau} = \left({\prod_{i=1}^n}{\otimes} f(k_i)
\right)_{\sigma\tau}.
\end{equation}

Let us present the tensor product ${\cal E}$ of Eq.(\ref{azimut}) as
follows:
\begin{equation}\label{decompos}
{\cal E}_{\sigma\rho}^{(n)} =  g_{\sigma\rho} - \sum_{i=1}^n
2\frac{(k_i)_{\sigma} (k_i)_{\rho}}{k_i^2} + \sum_{i>j}^n
2\frac{(k_i)_{\sigma} 2(k_i k_j)(k_j)_{\rho}}{k_i^2 k_j^2} + \ldots
\end{equation}

As soon as in DLA $k_i^{\perp 2}\simeq k_i^2$, the azimuthal
integration of the first two terms of the tensor  ${\cal E}$ results in
\begin{equation}\label{eresult}
E^{(n)}_{\sigma\tau} =
\int_{0}^{2\pi}\!\frac{d\phi_1}{2\pi}
\ldots
\int_{0}^{2\pi}\!\frac{d\phi_n}{2\pi}
{\cal E}^{(n)}_{\sigma\tau}
\simeq g_{\sigma\tau} - n g_{\sigma\tau}^{\perp}
\end{equation}
where $n$ is the number of quark cells in the ladder graph.

The analysis in\cite{ET2157} shows that the next terms of
the expansion Eq.(\ref{decompos}) after integration over azimuthal
angles in DLA contribute nothing.

Using the simple relation
$$
\frac{\partial\alpha_s^n}{\partial\ln\alpha_s}  = n \alpha_s^n
$$
one can rewrite the antisymmetric hadronic tensor Eq.(\ref{DLladd})
as
\begin{equation}\label{result}
W_{[\mu\nu]}^{(n)} = i\varepsilon_{\mu\nu\lambda\sigma}
\frac{mq^{\lambda}}{(pq)} \left\{ M^{(n)}{\cal S}^{\sigma} -
\frac{\partial}{\partial\ln\alpha_s} M^{(n)} {\cal
S}_{\perp}^{\sigma} \right\}
\end{equation}
with
\begin{equation}\label{moper}
M^{(n)} = \left(\frac{\alpha_s C_F}{2\pi}\right)^n \prod_{i=1}^n
\left[ \int_{x}^{\beta_{i-1}}\frac{d\beta_i}{\beta_i}
\int_{\kappa_i}^{\beta_i s}\frac{dk_{i\perp}^2}{k_{i\perp}^2} \right]
\beta_n \frac12 e_q^2 \delta(\beta_n\!-\!x\!-\!\frac{k_{n\perp}^2}{s}).
\end{equation}

Summing up over all ladders thus leads to a very simple relation
between the contributions to DIS structure functions $g_1^{NS}$ and
$g_2^{NS}$:
\begin{equation}\label{ladreltn}
g_2^{NS} = - \frac{\partial g_1^{NS}}{\partial\ln\alpha_s}
\end{equation}

Eq.(\ref{ladreltn}) is valid  only for ladder Feynman graphs.
To understand the effect of nonladder
graphs on Eq.(\ref{ladreltn}), let us add a nonladder gluon to a
ladder graph. Nonladder gluon must be soft enough not to destroy the DL
pattern of a ladder graph.  Unlike a skeleton ladder gluon a nonladder
gluon invokes an additional DL integration only if no any $k_{\perp}^2$
factor appears in the integrand's numerator. Such factor would cancel
the softest  virtuality propagator (of those invoked by a nonladder
gluon) in the denominator.  Thus adding a nonladder gluon does not
change the relation between contributions of the ${\cal S}^{\perp}$ and
 ${\cal S}^{\|}$-structures.

Therefore we would use Eq.(\ref{ladreltn}) if one could separate
out the contribution of ladder skeleton gluons to $g_1^{NS}$ and
differentiate it w.r.t. $\alpha_s$. Fortunately one can use the
expression for $g_1^{NS}$ obtained in the work~\cite{BER1}  and
tag the ladder and nonladder gluon contributions through
definition of separate QCD couplings $\alpha_{L}$ and $\alpha_{NL}$
respectively, $\alpha_{L} = \alpha_{NL} = \alpha_s$.~\cite{ET2157}

Then Eq.(\ref{ladreltn}) can be presented in the form
\begin{equation} \label{finalres}
g_2^{NS} = -\left.\frac{\partial
g_1^{NS}}{\partial\ln\alpha_L}\right|_{\alpha_L = \alpha_{NL}  =
\alpha_s}
\end{equation}
that predicts  the same asymptotics of $g_1^{NS}$ and  $g_2^{NS}$  at
$x\to 0$:
\begin{equation} \label{as}
g_2^{NS} \sim g_1^{NS}\sim \left( \frac{1}{x}\right)^a
\left(\frac{Q^2}{\mu^2} \right)^{\frac{a}{2}}, \qquad
a\simeq\sqrt{\frac{2\alpha_s C_F}{\pi}}
\left( 1\!+\!\frac{1}{2N_c^2} \right) .
\end{equation}

We have considered above the nonsinglet contribution to $g_2$.
Obviously, one should expect the singlet contribution,
i.e. insertion of gluon ladders, to be dominating over the nonsinglet
contribution~\cite{BER2}.
One could repeat the
analysis presented above for the case of gluon ladder contributions,
the crucial points of the analysis as well as general features of the
result remain the same. The only difference stems from different
contributions of a gluon rung to the numerator's tensor structure
Eq.(\ref{split}) for each case: for a quark ladder - Eq.(\ref{ftensor})
and for a gluon ladder -
\begin{equation}\label{htensor}
h^{\sigma\sigma^{\prime}}(k) = g^{\sigma\sigma^{\prime}} -
\frac{k^{\sigma}k^{\sigma^{\prime}}}{k^2}.
\end{equation}

Thus we obtain the following generalization of the expression
(\ref{finalres}) for the total structure function $g_2^S$:
\begin{equation}\label{totg2s}
g_2^S = -\left.\frac{\partial
g_1^S}{\partial\ln\alpha_L}\right|_{\alpha_L = \alpha_{NL} =
\alpha_s} -\ \frac12\left.\frac{\partial
g_1^S}{\partial\ln\widetilde{\alpha}_L}\right|_{\widetilde{\alpha}_L =
\alpha_{L} = \alpha_{NL} = \alpha_s}.
\end{equation}
where we differ the QCD coupling $\widetilde{\alpha}_L$ corresponding
to gluon radiation in a gluon cell from the QCD coupling
$\alpha_L$
corresponding to gluon radiation in a quark cell of a ladder graph.

The analytical expression\footnote{In the
previous work \cite{BER2} the small-$x$ asymptotics of $g_1^S$ was
calculated numerically.} for $g_1^S$ at small $x$ is
\begin{equation}\label{g1asan}
g_1^S=\frac{\sum e_q^2}{4} \int\frac{d\omega}{2\pi
i}\left(\frac1x\right)^\omega \frac{\omega
\left(2\omega\!+\!r_1\!+\!r_2\!+\!\frac{b_{(-)}}{h}(r_1\!-\!r_2)\right)
\exp(\Omega y)}{\omega^2+\omega (r_1\!+\!r_2) + r_1 r_2}
\end{equation}
where
\begin{eqnarray}
& &r_1=\sqrt{\omega^2\!-\!2b_{(+)}\!+\!2h},\qquad
r_2=\sqrt{\omega^2\!-\!2b_{(+)}\!-\!2h}, \nonumber \\
& &h=\sqrt{b_{(-)}^2\!+\!4b_1b_2}, \qquad
\Omega=\frac{\omega-r_2}{2}, \qquad y=\ln\frac{Q^2}{\mu^2}, \nonumber
\end{eqnarray}
and
\begin{eqnarray}\label{bs}
& &b_{(\pm)}=\frac{\alpha_L}{2\pi}C_F \pm
\frac{\widetilde{\alpha}_L}{2\pi}4N_c +
\left(\frac{\alpha_{NL}}{2\pi}\right)^2
\left(\frac{2C_F}{N_c}\mp 8N_c^2\right)\frac{1}{\omega^2}, \\
& &b_1=\frac{\alpha_L}{2\pi}2C_F -
\left(\frac{\alpha_{NL}}{2\pi}\right)^2 \frac{4C_FN_c}{\omega^2},
\qquad b_2=-\frac{\widetilde{\alpha}_L}{2\pi} +
\left(\frac{\alpha_{NL}}{2\pi}\right)^2 \frac{2N_c}{\omega^2}
\end{eqnarray}
where $1/\omega^2$-terms correspond to nonladder gluon contributions
so that color octet amplitudes are taken in the Born approximation. This
significantly simplifies formulae but does not noticably change the
small-$x$ asymptotics of $g_1^S$.

Thus asymptotics of $g_2^S$ and $g_1^S$ at $x\to 0$ is
\begin{equation} \label{ass}
g_2^{S} \sim g_1^{S}\sim \left(\frac1x\right)^{\omega_0}
\left(\frac{Q^2}{\mu^2}\right)^{\frac{\omega_0}{2}}
\end{equation}
where $\omega_0$ is the rightmost singularity (branching point) of the
integrand in Eq.(\ref{g1asan}):

\begin{equation}\label{singul0}
\omega_0=2b_{(+)}(\omega_0)+2h(\omega_0).
\end{equation}

This is an algebraic equation of the fourth order. Solution of this
equation gives
$\omega_0\simeq 3.5\sqrt{\alpha_s/2\pi}$.  If one neglects quark
contributions to the ladder then
$\omega_0\simeq 3.7\sqrt{\alpha_s/2\pi}$.  The small difference between
these results shows that similar to Pomeron gluons give the dominating
contribution to asymptotics of $g_2^S$ and $g_1^S$.

Let us recall that the above results were obtained in the DLA.
Worthwile to notice that taking single log effects into account,
though does not change the power-like character of the
Eqs.(\ref{as},\ref{ass}), changes the values of the powers.
For $g_1^{NS}$ the new power is
\begin{equation} \label{sl}
a^{\prime} = a \left[\left(1+\frac{a^2}{16}\right)^{\frac12}
+ \frac{a}{4}\right].
\end{equation}
For example, if one takes $\alpha_s=0.18$, the Eq.(\ref{sl}) gives:
$a\simeq 0.4$, $a^{\prime}\simeq 0.44$.


\begin{thebibliography}{99}
\bibitem{BER1} J.Bartels, B.I.Ermolaev and M.G.Ryskin,
Z.Phys.C 70 (1996) 273.
\bibitem{BER2} J.Bartels, B.I.Ermolaev and M.G.Ryskin,
Z.Phys.C 72 (1996) 627.
\bibitem{ET2157} B.I.Ermolaev and S.I.Troyan, Preprint PNPI TH-14-1997
2157;\\ hep-ph/9703384
\end{thebibliography}
\end{document}